\documentstyle[aps,epsfig,twocolumn,prl]{revtex}

\begin{document}
\draft
\author{Zbyszek P. Karkuszewski, Christopher Jarzynski and Wojciech H. Zurek}
\address{
Los Alamos National Laboratory, Theoretical Division, Los Alamos, 
New Mexico 87545, USA } 
\title{Quantum Chaotic Environments, the Butterfly Effect, and Decoherence}

\maketitle

\begin{abstract}
We investigate the sensitivity of quantum systems
that are chaotic in a classical limit, to small perturbations of their 
equations of motion.
This sensitivity, originally studied in the context of defining quantum chaos,
is relevant to decoherence in situations when the environment has a chaotic
classical counterpart.

\end{abstract}

The exponential divergence of two trajectories, evolving under identical 
equations of motion from slightly different initial conditions -- the famous 
butterfly effect -- is a fingerprint of chaos in classical mechanics. 
However, an analogous definition of ``quantum chaos''
based on evolution in Hilbert space is problematic:
by unitarity, the overlap
between two evolving wavefunctions
-- a natural indicator of distance between them --
is preserved with time, hence there is no divergence.
To address this difficulty, Peres \cite{AP} has suggested
an alternative approach,
in which one considers two trajectories evolving
(in phase space or Hilbert space)
from identical initial conditions but under slightly
different equations of motion,
rather than the other way around.
Classically, even for small perturbations \cite{small},
one generically expects rapid divergence when
the systems are chaotic according to the usual definition,
as the perturbation (i.e.\ the difference between equations of motion)
soon introduces a small displacement between the trajectories.
Quantally, the overlap between the wavefunctions begins at unity,
then decays with time, and Peres suggested the rate of this decay
-- a measure of the sensitivity of quantum evolution to
perturbations in the equations of motion --
as a signature of quantum chaos.

The sensitivity of quantum evolution also plays an important
role in the context of {\it environment-induced decoherence}
\cite{Zurek91,Giulini}.
As an illustrative example, consider a composite system consisting of
a two-state {\it spin} ($s$) and a generic {\it environment} 
(${\cal E}$), governed by a Hamiltonian 
of the form
\begin{equation}
\hat{\cal H} = 
\hat H_{\cal E} \otimes I_s \,+\,
\hat V_{\cal E}^+ \otimes \vert+\rangle\langle+\vert_s \,+\,
\hat V_{\cal E}^- \otimes \vert-\rangle\langle-\vert_s.
\end{equation}
Here the identity $I_s$ and the two projection operators
act on the Hilbert space of the spin, whereas
$\hat V_{\cal E}^\pm$ and $\hat H_{\cal E}$ act on that
of the environment.
We view $\hat H_{\cal E}$ as the ``bare'' Hamiltonian for
the environment, and $\hat V_{\cal E}^\pm$ as a perturbative
coupling to the state of the spin.
An initial state $\Psi(0) = 
\psi_{\cal E}(0) \Bigl[\alpha\vert+\rangle_s + \beta\vert-\rangle_s\Bigr]$
evolves into
\begin{equation}
\Psi(t) = \alpha\psi_{\cal E}^+(t)\vert+\rangle_s
+ \beta\psi_{\cal E}^-(t)\vert-\rangle_s,
\end{equation}
where the unitary evolution of $\psi_{\cal E}^\pm(t)$
in the Hilbert space of the environment
is generated by $\hat H_{\cal E}+\hat V_{\cal E}^\pm$.
The initially pure state of the spin, 
$\alpha\vert+\rangle_s + \beta\vert-\rangle_s$,
eventually becomes a mixture of the pointer states 
$\{|+\rangle_s, |-\rangle_s \}$ as a result of monitoring by the environment.
The decay of $\vert\langle\psi_{\cal E}^-\vert\psi_{\cal E}^+\rangle\vert^2$
is an indicator of this process:
once this overlap becomes negligible, the state of the spin alone can be
described in terms of classical probabilities
rather than quantum amplitudes.

In view of these considerations, we are motivated
to ask,
{\it what limits are placed on the sensitivity of a
quantum system to perturbations in its equations
of motion?}
The aim of this Letter is to provide answers to
this question, with emphasis on systems that are
chaotic in the classical limit.
The object of our considerations will be a pair
of wavefunctions, $\psi_1(t)$ and $\psi_2(t)$,
identical at $t=0$, that evolve under slightly
different Hamiltonians,
$\hat H_1$ and
$\hat H_2\equiv \hat H_1 + \hat V$,
respectively.
Our measure of ``sensitivity'' will be the rate
of decay of the overlap
\begin{equation}
\label{eq:oq}
O_q(t) =
\vert \langle \psi_1(t)\vert\psi_2(t) \rangle \vert^2.
\end{equation}
We will first derive, from the uncertainty principle,
a quite general bound on the rate of this decay.
We will then clarify the difference in robustness
between classical chaotic systems and their
quantum counterparts, in terms of the size of structures
found in corresponding phase space functions
(classical probability distributions and quantum Wigner functions).
Finally, we will illustrate the central issues
with a numerical example, placing
bounds on the time needed for a
(classically chaotic) quantum environment to decohere
a quantum system of interest.

{\bf Quantum lower bound for overlap decay and decoherence time.}
Using the projection operator
$\hat P(t) = \vert\psi_2\rangle\langle\psi_2\vert$
to rewrite the above-defined overlap as
$O_q(t) = \langle\psi_1\vert\hat P\vert\psi_1\rangle$,
and applying the Schr\" odinger equation in the
Heisenberg picture, we obtain
\begin{equation}
\label{eq:cj_doqdt}
{d O_q\over dt} =
-{i\over\hbar}
\langle
[\hat V,\hat P]\rangle,
\end{equation}
where $\langle\cdots\rangle\equiv
\langle\psi_1\vert\cdots\vert\psi_1\rangle$.
Now, the uncertainty relation for
$\hat V$ and $\hat P$ is
$\Delta V \Delta P \ge
\vert\langle
[\hat V,\hat P]\rangle\vert/2$,
where
$(\Delta V)^2 =
\langle\hat V^2\rangle - \langle\hat V\rangle^2$
is the variance of the operator $\hat V$ in the
state $\psi_1$,
and similarly
$(\Delta P)^2 = \langle\hat P^2\rangle - \langle\hat P\rangle^2
= O_q - O_q^2$.
Combining this with (\ref{eq:cj_doqdt}) gives
\begin{equation}
-{dO_q\over dt} \le \Bigl\vert{dO_q\over dt}\Bigr\vert\le
{2\over\hbar}\Delta V (O_q-O_q^2)^{1/2},
\end{equation}
leading, after some algebra, to the inequality
\begin{eqnarray}
O_q(t) \ge \cos^2\left(\frac{1}{\hbar}\int\limits_0^t\Delta V\mbox{d}t'\right)
\equiv \cos^2 \phi(t),
\label{LB}
\end{eqnarray}
valid until $\phi(t)$ 
(which never decreases) reaches $\pi/2$.
Note that by reversing the roles of $\psi_1$
and $\psi_2$ in this argument, we typically obtain a quantitatively
different, though equally valid, result.
We can therefore view $\Delta V$ appearing in
(\ref{LB}) as the spread of $\hat V$ in
either state $\psi_1$ or $\psi_2$, whichever gives
the tighter bound.

When $\psi_1$ and $\psi_2$ represent states of
a quantum environment (as discussed above), then (\ref{LB}) gives
the following lower bound on the decoherence times:
\begin{equation}
\tau_D \agt \pi\hbar/2\overline{\Delta V},
\end{equation}
where $\overline{\Delta V}$ is to be interpreted as the
typical value of $\Delta V$ during the decoherence process.

{\bf Quantum and classical overlap in terms of phase space distributions.} 
Apart from studying the sensitivity of quantum evolution
in its own right, we would like to compare it with classical
sensitivity, particularly in the case of chaotic evolution.
We will work with functions in phase space
as these transparently suggest a classical counterpart
of the quantum overlap $O_q(t)$.

Equation \ref{eq:oq} can be rewritten as\cite{1dof}
\begin{equation}
\label{eq:oqwig}
O_q(t) = 2\pi\hbar \int W_1(x,p,t)W_2(x,p,t)~\mbox{d}x~\mbox{d}p,
\end{equation}
where the $W_i$'s are Wigner functions corresponding
to $\psi_1$ and $\psi_2$,
evolving under $\hat H_1$ and $\hat H_2$.
Let us now consider two classical phase space distributions,
$L_1(x,p,t)$ and $L_2(x,p,t)$, obeying the Liouville
equation under the respective classical Hamiltonians
$H_1(x,p)$ and $H_2(x,p)$.
Let us furthermore set the initial conditions for the
$L$'s to be the same as those for the $W$'s:
$L_1 = L_2 = W_1 = W_2$ at $t=0$.
In view of (\ref{eq:oqwig}) it is now natural to
define a classical overlap,
\begin{equation}
\label{eq:co}
O_c(t) = 2\pi\hbar \int L_1(x,p,t) L_2(x,p,t)~\mbox{d}x~\mbox{d}p,
\end{equation}
where the (arbitrary) normalization factor $2\pi\hbar$
was chosen so that $O_c(0) = O_q(0) = 1$.
By comparing the decay times of $O_q(t)$ and
$O_c(t)$, we now have a setup for comparing
quantum and classical sensitivity to perturbations
in the equations of motion.
Because of the relevance to decoherence, we will refer to these 
as the quantum and classical {\it decoherence times},
though in the classical case this is just convenient nomenclature.

{\bf The smallest structures of phase space distributions}.
A central hypothesis of this Letter is that the time scale
for the decay of the overlaps $O_c$ and $O_q$ is determined primarily 
by the size of the smallest structures in the corresponding
phase space distributions, with particular relevance
when the classical evolution is chaotic.
In both cases we have
two initially identical phase space distributions (the $L$'s or $W$'s)
evolving with time while slowly accumulating a relative
displacement due to the perturbation;
a substantial decay of overlap occurs when this
displacement is large enough that the two functions
no longer ``sit one on top of the other''.
Clearly, this depends not only on the rate at which
the functions move apart, but also on the local
smallness of their structure, as this determines
the degree of displacement needed to kill the overlap.
The difference between the decay of overlap in the
classical, chaotic case, and in its quantum counterpart,
arises because of the qualitatively different mechanisms
governing the emergence of small-scale details
in the corresponding phase space distribution.

In the classical case, the size of local structure in
$L_1$ and $L_2$ shrinks exponentially with time, due to the
stretching and folding associated with chaotic evolution:
the probability distributions become thin and elongated,
with a local width decreasing as $\exp(-\lambda t)$,
where $\lambda$ is the largest Lyapunov exponent.
As there is no lower bound to this smallness, it is
clear that the decay time will be set predominantly
by the Lyapunov time. By contrast, there are limits on the fineness of detail
that can develop in the Wigner function, $W$;
e.g.\ the Wigner function of a superposition of two 
identical Gaussians separated by $\Delta X$ -- a
Schr\" odinger cat-like state -- exhibits interference fringes 
in momentum on a scale $\delta p \simeq \hbar/\Delta X$ \cite{Zurek91}.  
More generally, when
spread over an area $A = \Delta X \Delta P$ in two-dimensional phase 
space, $W$ exhibits local structure on scales $\delta p \simeq \hbar / \Delta X$,
$\delta x \simeq \hbar / \Delta P$ \cite{Habib,Nature}. The corresponding
phase space scale is associated with the sub-Planck action $a \sim \hbar^2/A$ 
which has physical consequences \cite{Nature}.
Most notably (in the present context),
the decay of $O_q$ occurs when the relative displacement of $W_1$ and $W_2$
is sufficient for their respective
smallest-scale fringes to interfere destructively.

Two examples serve to build up intuition related to these issues.

Example I. 
Let us assume that identical wave functions $\psi_1$ and $\psi_2$ are 
superpositions of $N$ Gaussians 
$\tilde G_j\sim \exp(-(x-x_j)^2/(2\hbar))\exp(ip_jx\hbar)$ 
$$
\psi_1(x)=\psi_2(x)=\sum_{j=1}^N \tilde G_j(x;x_j,p_j).
$$ 
The corresponding Wigner functions $W_1$ and $W_2$ consist of
$N$ coherent-state Gaussians
$G_j$, centered at points $(x_j,p_j)$,
as well as pairwise interference terms $G_{j,k}$:
$$
W=\sum_{j=1}^N G_j + \sum_{j<k}G_{j,k}.
$$
We assume (following \cite{Nature}) that the coherent-state Gaussians
are sparse: each pair $G_j$ and $G_k$ is well separated by the
``distance'' $d_{j,k}$ phase space.
The interference term $G_{j,k}$ is then another Gaussian located halfway
between $G_j$ and $G_k$,
modulated by an oscillatory factor of frequency
$d_{j,k}/\hbar$ and twice the amplitude 
of $G_j$ \cite{Zurek91}. 
The overlap (\ref{eq:oqwig}) of $W_1$ and $W_2$ then works
out to be $O_q\approx Ng + (N-1)Ng$, 
where $g=\int G_j^2 \mbox{d}x\mbox{d}p$. 
The first term corresponds to the
overlap deposited in the coherent Gaussians $G_j$,
the second in the interference terms $G_{j,k}$.
Thus, for large $N$ {\it most of the overlap resides in the interference terms}.
If we now displace one of the WF's relative to the other,
by a distance at least on the order of the size of a typical interference fringe
(but small compared to the size of the $G_j$'s),
then the contributions to the overlap from the $G_{j,k}$'s will
typically interfere destructively, resulting in a total overlap
$\sim 1/N \ll 1$.
This is somewhat counterintuitive:
for a fixed number of Gaussians of fixed size,
we can increase the sensitivity of the system
-- as measured by the perturbation needed to 
kill the overlap $\vert\langle\psi_1\vert\psi_2\rangle\vert^2$ --
simply by increasing the average distance between the Gaussians,
or equivalently the total area occupied in phase space.

Example II.
While classical probability distributions have no interference
fringes, there is no bound on the smallness of
structures resulting from chaotic stretching and folding.
Let us examine a simplified model of the classical evolution
of two Gaussian distributions in phase space, initially identical:
$$
L_1(x,p,t=0)=L_2(x,p,t=0)=\frac{1}{2\pi\sigma_x\sigma_p}
e^{-\frac{x^2}{2\sigma_x^2}-\frac{p^2}{2\sigma_p^2}},
$$
with $x$ and $p$ dimensionless.
We now assume that with time both $L$'s are
exponentially stretched in the $p$ direction and squeezed in the
$x$ direction, in an area-preserving way:
$\sigma_p(t)=\sigma\exp(\lambda t)$,
$\sigma_x(t)=\sigma\exp(-\lambda t)$;
furthermore, while the centroid of $L_1$ remains fixed
at $x=p=0$, the centroid of $L_2$ drifts
with a constant velocity ${\bf v} = (v_x,v_p)$.
These assumptions mock up the relevant features of chaotic evolution
under slightly different Hamiltonians,
where $\vert{\bf v}\vert$ is an indicator of the size of the
perturbation.
A simple calculation gives us the following decay of the overlap
between $L_1$ and $L_2$:
\begin{equation}
O_c(t) = 
e^{-(v_x t \cdot e^{\lambda t}/2\sigma)^2}
e^{-(v_p t \cdot e^{-\lambda t}/2\sigma)^2}.
\label{eq:co2}
\end{equation}
Generically, the first factor will dominate, and the overlap
will decay to negligible values on a time scale set
by $\lambda^{-1}$.

While both of these examples are highly simplified, we believe they
capture the essential physics.
We now present numerical results illustrating actual evolution.

\begin{figure}[htb]
\centering
{\epsfig{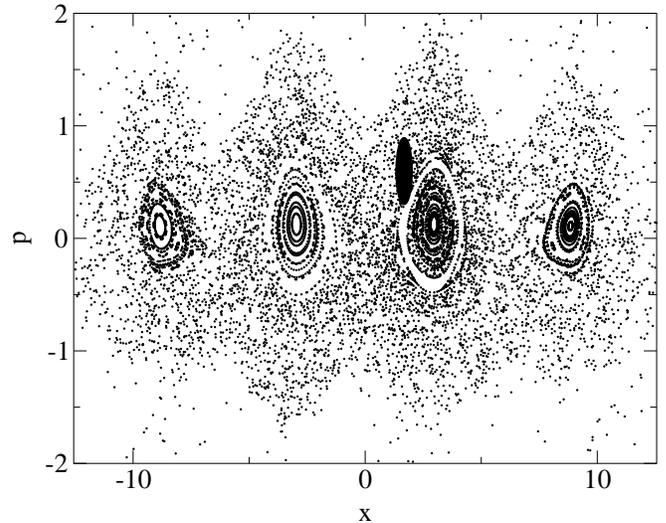}}
\caption{Poincar\'e surface of section. The black ellipse shows the initial
distribution. The evolution is confined to the dark cloud surrounding
the islands of stability.} 
\label{fig1}
\end{figure}

{\bf Numerical simulation}. 
An example of a time-dependent Hamiltonian that generates chaos
in one dimension is\cite{ehrenfest}:
\begin{equation}
H=\frac{p^2}{2m} - \kappa\cos(x-l\sin(t)) + a\frac{x^2}{2}.
\end{equation}
For parameter values $m=1,\>\kappa=0.36,\>l=3.8$ and $a=0.01$, the
stroboscopic Poincar\'e surface of section, Fig.\ref{fig1}, consists of 
four islands of stability surrounded by a chaotic sea. 
We have simulated both quantum and classical evolution, starting from
a Gaussian distribution centered just outside the regular region
(see Fig.\ref{fig1}) at $t=0$, and evolving under $H$ until time $t=T$,
at which point a perturbation is turned on and the evolution forks into
two branches governed by the Hamiltonians
\begin{equation}
H_\pm = \frac{p^2}{2m} - \kappa\cos(x-l\sin(T+\tau)) + a\frac{(x\pm\epsilon)^2}{2},
\end{equation}
where $\epsilon=0.5$ and $\tau\equiv t-T$. 
The perturbation is thus
$V = H_+-H_- = 2 a \epsilon x$.
The preparation time interval ($0\le t\le T$)
allows the distributions to develop small
structures in phase space.
After the perturbation is turned on at $\tau=0$, we
monitor the decay of overlaps.

\begin{figure}[htb]
\centering
{\epsfig{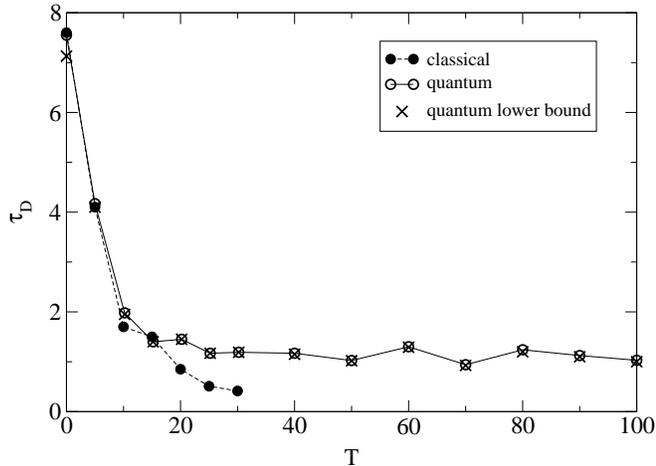}}
\caption{Time after which the overlap decreases to 0.9 versus preparation time for classical and quantum
environments.} \label{fig2}
\end{figure}

Figure \ref{fig2} shows how the decoherence time $\tau_D$ (defined here
somewhat arbitrarily as
the time $\tau$ at which the overlap decreases to a value 0.9\cite{explain10percentdecay})
depends on preparation time $T$ for quantum and classical evolution.
For short preparation times, both are equally sensitive
to the perturbation applied at $\tau=0$,
reflecting the fact that the size of the smallest structure is
basically the same in both cases.
However, once the distributions have spread over much of the
dynamically accessible area in the phase space, which occurs at 
$T\approx20$, the size of interference fringes in the Wigner functions
saturates, resulting in more or less
constant decoherence times even for long $T$.
Note that the
quantum lower bound (\ref{LB}) denoted by crosses in the Fig.\ref{fig2},
with $\Delta V$ evaluated directly from the simulation,
gives results very close to the actual decoherence times;
this indicates that the quantum states used in our evolution are
close to minimum uncertainty states with respect to the
uncertainty principle mentioned after (\ref{eq:cj_doqdt}).
In the absence of such structure saturation in the classical distributions $L$,
the classical decoherence time continues to decrease with increasing
preparation time, due to the presence of ever smaller structures.
While the computational cost of the classical simulations became prohibitive
for $T>30$, the observed decrease in $\tau_D$ is consistent with
a rapid approach toward zero.

To further investigate the relevance of the smallest
structure scale in the Wigner function to the decay of $O_q$,
let us define the spread of the system in position as
$\Delta X=(\langle \hat x^2\rangle - \langle \hat x \rangle^2)^{1/2}$,  and 
in momentum
$\Delta P=(\langle \hat p^2\rangle - \langle \hat p \rangle^2)^{1/2}$, with
averages $\langle \cdots \rangle$ taken at $\tau=0$.
This translates to interference fringes of size
$\delta x\approx\hbar/{\Delta P}$ in position and
$\delta p\approx\hbar/{\Delta X}$ in momentum.
Using our data,
we have determined $\delta x$ and $\delta p$ for each of the
fourteen preparation times $T$ shown in Fig.\ref{fig2}.
\begin{figure}[htb]
\centering
{\epsfig{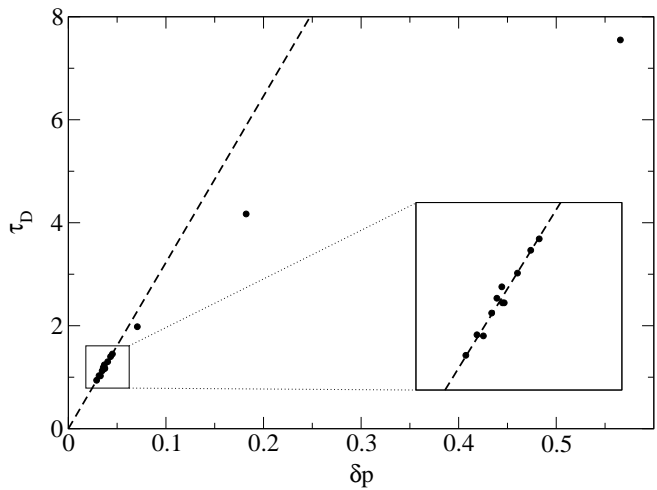}}
\caption{Time after which overlap falls by 10\% versus $\delta p$.} \label{fig3}
\end{figure}
For the parameter values and form of the perturbation
we have chosen, we have found that
$\delta p$ is more relevant than $\delta x$ for the decay of $O_q(t)$.
Therefore in Fig.\ref{fig3} we plot the dependence of $\tau_D$ on $\delta p$.
The linear dependence observed for small values of $\delta p$ (large $T$)
is not unexpected.
Recall that decoherence is achieved when
the displacement of $W_2$ relative to $W_1$ becomes comparable
to $\delta p$.
In the regime of small $\delta p$, this occurs
sufficiently rapidly that the value
of $\delta p$ does not change much during the process.
Hence, given a constant rate of relative drift in the momentum direction
(due to the form of our perturbation, $\hat V = 2a\epsilon\hat x$),
we expect $\tau_D\propto \delta p$.
On the other hand, when $\delta p$ is initially large (small $T$),
then the decoherence time will also be large, and
$\delta p$ itself will decrease during this time;
hence, we expect in this regime to obtain values of $\tau_D$
that are smaller than suggested by the initial value of
$\delta p$, in agreement with the three highest data points
shown in Fig.\ref{fig3}.

{\bf Conclusions}.  We have investigated the sensitivity 
of classical chaotic systems and their quantum counterparts to perturbations 
in their equations of motion.
From general quantum considerations, we have derived a lower
bound for the decay of $O_q(t)$.
We have further argued that the sensitivity (in both the classical
and quantum cases) is set by the size of the smallest structure
of the related phase space functions.
Finally, we have discussed the relevance of these results to
the ability of quantum environments to rapidly decohere systems
to which they are coupled.

We are grateful to Diego Dalvit and Salman Habib for comments
on the manuscript and stimulating discussions.
All calculations presented here were performed on the
Avalon Beowulf cluster at Los Alamos National Laboratory.
This research is supported by the Department of Energy,
under contract W-7405-ENG-36.


\begin{references}
\bibitem{small} 
By ``small'' we mean that the spectra of Lyapunov exponents
for the two Hamiltonians are essentially the same.
In particular, this implies that the perturbation neither
breaks nor introduces symmetry.
The hierarchy of small perturbations based on energy level spacing 
is discussed in \cite{Loschmidt}
\bibitem{1dof}
For simplicity we will work in one degree of freedom, but the
results extend easily to higher dimensions.
\bibitem{explain10percentdecay}
This threshold, while admittedly somewhat close to unity,
was motivated by computational necessity:
with a lower threshold we would not have been able to
obtain reliable classical estimates of $\tau_D$ for all
of the points (up to $T=30$) shown in Fig.\ref{fig2}.
\end{references}
\end{document}